# Bayesian method for evaluation an airline's profitability on the base components of Airline Route Planning

L. V. Androshchuk and Aidan Rooney

Airline route planning takes into account the factors of commercial and customer preferences, safety, and should allow a flexibility given the tremendous uncertainty about market conditions. The mathematical model on the Bayes formula allows optimizing the interaction of the base factors.

The modelling results applied in network planning allow an airline management team to understand what is required, to provide an overview of the airline planning process, from the longest-range strategic decisions involving aircraft acquisition to medium-term decisions related to route planning and scheduling.

Large amounts of real time and historical data must be addressed, estimated, and taken into account to establish an appropriate mathematical model for prediction and evaluation influence of airline route planning on an airline's profitability. The best results can be achieved through a balance of all components at the same time. In reality it is not easy to take into account all variables, but taking into consideration the main areas, significant economic effect can be obtained.

The most important planning decisions faced by airline management can be categorized as follows:
- route planning establishes the geography of profitable flights, subject to fleet availability constraints;
- fleet planning determines the types, quantity of aircraft to acquire, and when to purchase them;
- schedule development is charged with operating frequency and time of the flights on each route, subject to operational and aircraft limitations.

Fleet Planning Methods are mainly based on statistical data. Airline decisions related to fleet planning depend primarily on an evaluation of the expected impacts of new aircraft on the airline's economic and financial performance:

- forecast of expected traffic;
- target average load factor;
- the number of aircraft required.

Fleet Planning Evaluation use two methods. A "top-down" or "macro" approach based on relatively high-level aggregate analysis, and a "bottom-up" or "micro" approach based on much more detailed analysis of data and forecasts of flights and routes.

The process of route planning and evaluation involves the selection by the airline of which commercially viable routes should be flown and therefore what fleet, which aircraft, should be chosen. The route selection decision is both strategic and tactical. It is an essential component of an integrated network strategy or "vision" for the airline, which must decide whether to focus on short-haul or long-haul services, domestic or international operations. At the same time, the characteristics of the selected routes will affect the types of "products" the airline offers to travellers. For example, an international route network will likely lead to a decision that business and first-class products should be offered in order to be competitive.

Given the airline's choice of aircraft and a fleet plan that determines the availability of aircraft with different capacity and range characteristics, the next step in the airline planning process is to determine the specific routes to be flown. In some cases, the sequence of these decisions is reversed, in that the identification of a profitable route opportunity might require the acquisition of a new aircraft type not currently in the airline's fleet. Economic considerations and expected profitability drive route evaluations for most airlines.

Schedule Development determines the factors, which affect the volume of Origin-Destination / Hub & Spoke, air travel Demand. There are so many variable factors, which can and do affect the volume and therefore correct mix of origin – destination demand. These factors will be addressed to understand what components need to be evaluated and utilized in the mathematical models, which will be used to forecast / predict future trends in airline route planning to increase profit maximization for airlines.

The more advanced airlines and Revenue Management Systems, (RM Systems), vendors have developed computerized RM systems that can perform forecasting and optimization by booking class for each flight leg departure, in addition to having the same database and booking monitoring capabilities of previous systems. Based on a variety of empirical studies and simulation experiments performed by airlines and academics alike, it is now commonly accepted that proper use of a RM system can lead to airline revenue increases of 4 to 6% [1].

Beyond the obvious incremental revenue benefits, the use of RM systems allows for better tactical matching of demand vs. supply by the airline. With RM capabilities, an airline can match or initiate almost any low fare that covers variable passenger carrying costs. Computerized RM systems manage the airline's inventory of available seats by using mathematical models and computer databases to address three different problems:
- overbooking;
- fare class mix;
- Origin – Destination control.

There are well known methods to evaluate a company's profitability, and obviousely, fleet planing influences an airlines effeciency. Main logistics issues as structure and control can contribute reducing costs on tied up capital and can also have a positive effect on customer service, which, of course, also affects airline profitability. However, different components may have conflicting effect and different levels of importance for a company. Therefore, a question arises for optimisation of overall performance based on several variables. The Bayes method is apropriate to build the mathematical model in this case. Therefore, it needs coordinating revenues, a customer service, tied up capital, and costs.

Nowadays, the Bayesian methods [2] have become quite widespread and are actively used in various fields of activities. These methods have materialised with the development of information technology. When real data needs to be analyzed, it means some parameter of this data is interesting and must be calculated. Then the probabilistic value *P(Data)* of the parameter can be obtained on the basis of the available data.

Application of the law of total probability leads to the following:

$$P(Data) = \sum P(\theta)P(Data|\theta)$$

In fact, the total probability of an outcome *P(Data)* depends on several determined parameters, namely, hypotheses, that have different importance for a company. Since the company may vary a value of hypotheses, (in the other words the probability of hypotheses $P(\theta)$, the maximum probability of the outcome can be acheived. The likelihood function or the conditional probability *P(Data|θ)* is determined by the model of data collection and depends on the parameter of interest.

Besides, it is necessary to know the weight of each of the parameters in the outcome, when the probability of the outcome has been already estimated. The weights can be calculated with help of Bayes' theorem, which is stated as the following formula:

$$P(\theta|Data) = \frac{P(\theta)P(Data|\theta)}{P(Data)}$$

Probability *P(θ|Data)* is called a posteriori probability. In order to compute it the likelihood function *P(Data|θ)* and a priori probability *P(θ)* have to be known or estimated.

The quantity of hypotheses is detemined by the real performece conditions, and it is important, that the hypotheses represent a set of pairwise disjoint events whose union is the entire space of the sample. On the scheme below five hypotheses are figured: $H_1, H_2, ...H_5$, with the coresponding weights $P(H_1), P(H_2), ..., P(H_5)$. Since, the outcome *A* depends on the hypotheses weights, the conditional probabilities $P(A/H_1), P(A/H_2), ..., P(A/H_5)$ should be prespecified.

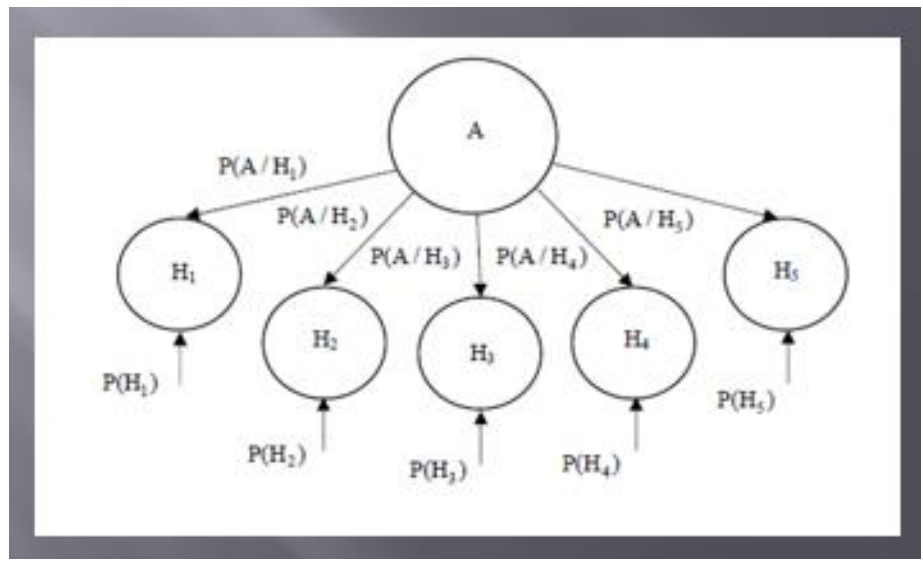

In the case of evaluation of an airline's profitability (*A*), it is important to balance three components:
- customer service,
- unavailable capital,
- costs,

which can be denoted as three hypotheses respectively. Then, corresponding to Bayes' theory, a level of importance for each of components $P(H_1), P(H_2), P(H_3)$, and the conditional probabilities of the profitability $P(A/H_1), P(A/H_2), P(A/H_3)$ can be denoted respectively (the scheme below).

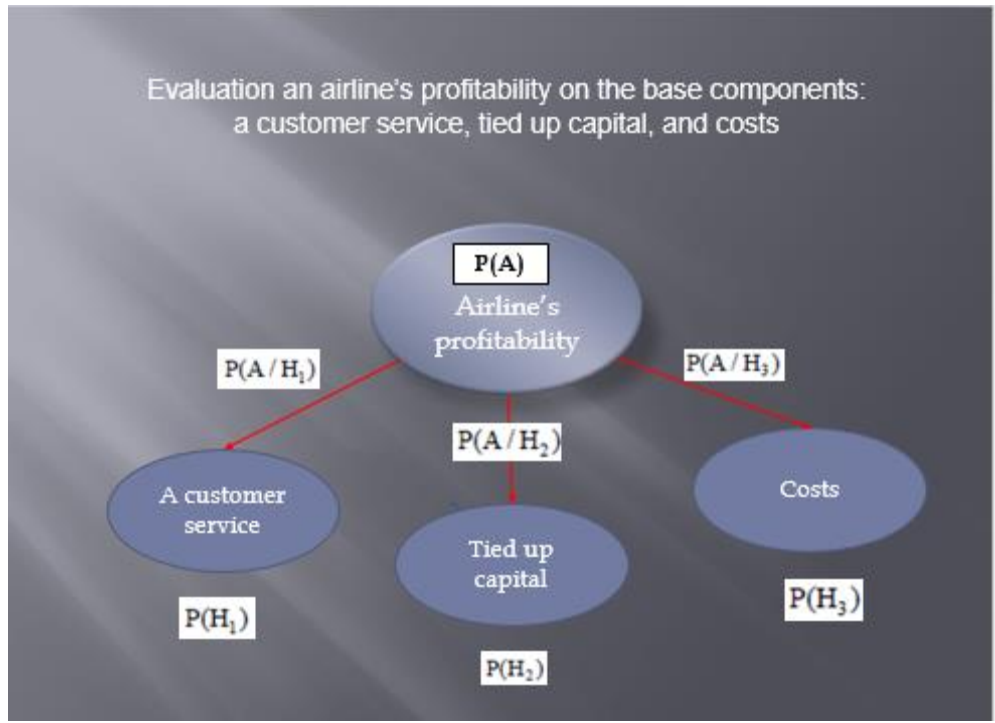

Thus, the the probability of predictable profit can be evaluated:

$$P(A) = P(H_1)P(A|H_1) + P(H_2)P(A|H_2) + P(H_3)P(A|H_3),$$

and by Bayes' theorem the contribution of each component into airline profitability is determined:

$$P(H_i|A) = \frac{P(H_i)P(A|H_i)}{P(A)}.$$

In accordance with Bayesian modelling, an appropriate control of the effect of certain components of the airline planning process on airline profitability can be maneged. Airline Route Planning is a system of components that can vary, when different players are involved like companies that establish business relations or service functions.

Coordinating these relations leads to a better total effect than taking into account some individual component. This is important, that in the Bayesian approach main elements can be balanced to get maximum economic result and provide customer preferences.